\documentclass[aps,prb,twocolumn,showpacs]{revtex4}
\usepackage{amsmath}
\usepackage{graphicx}
\newcommand{\al}{\alpha}
\newcommand{\be}{\beta}

\newcommand{\de}{\delta}
\newcommand{\e}{\epsilon}

\newcommand{\thi}{\theta}

\newcommand{\la}{\lambda}
\newcommand{\mi}{\mu}
\newcommand{\n}{\nu}
\newcommand{\p}{\pi}
\newcommand{\ro}{\rho}
\newcommand{\s}{\sigma}

\newcommand{\w}{\omega}
\newcommand{\G}{\Gamma}

\newcommand{\ta}{\tau}

\newcommand{\round}[1]{\left({#1}\right)}
\newcommand{\square}[1]{\left[{#1}\right]}

\newcommand{\ang}[1]{\langle #1\rangle}

\begin{document}

\title{Molecular transistor coupled to phonons and Luttinger-liquid leads}

\author{So Takei$^{1}$}
\author{Yong Baek Kim$^{1,2}$}
\author{Aditi Mitra$^{3}$}
\affiliation{$^{1}$Department of Physics, University of Toronto, Toronto,
Ontario M5S 1A7, Canada\\
$^{2}$School of Physics, Korea Institute for Advanced Study, Seoul 130-722,
Korea\\
$^{3}$Department of Physics, Columbia University, New York, 
New York 10027}
\date{\today}

\begin{abstract}
We study the effects of electron-phonon interactions on the transport properties of
a molecular quantum dot coupled to two Luttinger-liquid leads. In particular, we 
investigate the effects on the steady state current and DC noise characteristics. 
We consider both equilibrated and unequilibrated on-dot phonons.
The density matrix formalism is applied in the high temperature approximation
and the resulting semi-classical rate equation is numerically solved for 
various strengths of electron-electron interactions in the leads and 
electron-phonon coupling. The current and the noise
are in general smeared out and suppressed due to intralead electron 
interaction. 
On the other hand, the Fano factor, which measures the noise normalized 
by the current, is more enhanced as the intralead interaction becomes stronger.
As the electron-phonon coupling becomes greater than order one,
the Fano factor exhibits super-Poissonian behaviour. 
\end{abstract}


\maketitle

\section{Introduction}
In the relentless search for smaller electronic devices, 
the idea of fabricating extremely small transistors using quantum 
dots has become an important topic. 
In recent years, the possibility of using a very small molecule as the 
quantum dot in a transistor has become apparent\cite{smitetal02,
dattaetal97,porathetal97,reedetal97,bummetal96,pasupathyetal03,
jparketal02,grobisetal02,joachimetal00}. 
Such a device may be thought of as a molecular quantum dot 
weakly coupled to macroscopic charge reservoirs, or leads. 

Molecules integrated into these transistors often have very complex 
structures and therefore introduce new electronic transport 
properties. For example, the shape of the 
molecule or the position of the molecule relative to the leads
can change as charges are added onto the molecule. 
A number of experiments have
investigated the effects of coupling between tunneling electrons and the
on-dot vibrational quanta on various transport properties of the molecule
\cite{pasupathyetal03,zhitenevetal02,stipeetal98,hparketal00}. 
Of particular interest is the experimental evidence for electron-phonon
coupling in a molecular device composed of C$_{60}$ molecule deposited
between a pair of gold electrodes\cite{hparketal00}. Experimental results 
show peaks in the differential conductance which may be due to 
the effect of coupling between tunneling electrons and the molecular 
vibrational mode. 

Several theoretical works\cite{mitraetal04,braigetal03-1,
mccarthyetal03,flensberg03,braigetal03-2,ajietal03,lundinetal02,boeseetal01} 
have harnessed the study of electron-phonon coupling by introducing 
an on-dot phonon degree of freedom that couples to on-dot electrons and 
observed the effects the coupling has on various transport properties 
of the molecular transistor.
One of the most recent comprehensive work on phonon effects
in molecular transistors coupled to two {\it non-interacting} leads 
was conducted by Mitra {\it et al}.\cite{mitraetal04}. 
They developed the density matrix formalism to study 
transport properties for both in-equilibrium and out-of-equilibrium
phonon distributions. 

In our work, we investigate how the phonon distribution {\it and} electron-electron
interaction in the leads affect the steady state DC current and the DC noise
characteristics of a molecular transistor. The phonon distribution can 
drastically change depending on how they are allowed to relax 
as they couple onto a bath. As discussed in Mitra 
{\it et al}.\cite{mitraetal04}, we must consider two competing 
time scales: time scale for the phonons to reach equilibrium via their 
interaction with the bath, $\tau_{\rm ph}$; and the dwell time of the 
electron on the dot, $\tau_{\rm dwell}$.  
The two limiting cases considered in this work are: the phonons 
are equilibrated to the bath corresponding to $\tau_{\rm dwell} \gg \tau_{\rm ph}$; 
and the opposite case where the tunneling electrons see unequilibrated phonons
corresponding to $\tau_{\rm ph} \gg \tau_{\rm dwell}$. 

We are particularly interested in the effects of electron-electron interaction in the
leads. Up to this date, theoretical works have only considered non-interacting
electrons in the leads. This is sufficient if
the leads are two or three dimensional electron gases, where interactions
affect the low energy properties of the lead only perturbatively. However, in one
dimension, arbitrarily weak electron interaction completely changes the 
ground state. In our work, we consider the leads to be a Luttinger
liquid. Our work may represent a physical device 
composed of a molecular quantum dot, such as C$_{60}$, coupled to two metallic 
single-walled carbon nanotubes, which behave essentially as 
interacting one-dimensional objects.

The outline of this paper is as follows. We consider a molecule 
with a single level coupled to two Luttinger-liquid leads. We suppose that 
the electrons are coupled to the on-dot vibrational mode and consider 
both equilibrated and unequilibrated phonons. We solve the model in 
the density matrix formalism and apply the high-temperature approximation. 
The approximation will be used to derive the rate equation for the 
dot occupation probabilities and use the equation to compute the 
steady state DC current and the DC noise characteristics of the 
molecular transistor as a function of source-drain voltage.
In section \ref{formulation}, we introduce our model, provide a
brief outline of the density matrix formalism, and present expressions for
the tunneling current and the DC noise.
In section \ref{results}, we present numerical results for the current
and the noise and discuss the dependence of these transport properties on
the phonon distribution and the intralead electron interaction. 
We conclude in section \ref{conclusion}.

\section{Formulation of the problem}
\label{formulation}
\subsection{Model}
\label{model}
Our model considers a molecule with a single non-degenerate energy level,
of energy $\e_0$,
coupled to two Luttinger-liquid leads labelled $L$ (left) and $R$ (right). 
We allow the 
tunneling electrons to couple to a single phonon mode, with frequency 
$\w_0$ corresponding to the internal vibrational mode of the molecule.
The full Hamiltonian is given by
\begin{equation}
H=H_{dot}+H_{leads}+H_t,
\end{equation}
where $H_{dot}$ describes the on-dot electrons coupled to the single
phonon mode, $H_{leads}$ represents the electrons in the leads,
and $H_t$ corresponds to the tunneling between the dot and the leads.

More specifically, $H_{dot}$ has the form,
\begin{multline}
H_{dot}=\e_0\sum_\s c^{\dagger}_{\s}c_{\s}+\frac U2\sum_\s c^{\dagger}_{\s}c_{\s}
\round{\sum_\s c^{\dagger}_{\s}c_{\s}-1} \\ +\la\w_0\round{b^{\dagger}+b}
\sum_\s c^{\dagger}_{\s}c_{\s}+\w_0b^{\dagger}b,
\end{multline}
where $c^{\dagger}_{\sigma}$ is the creation operator of electrons with spin
$\sigma$ on the dot and $b^{\dagger}$ is the creation operator of the on-dot
phonons. $\lambda$ is the electron-phonon coupling strength and $U$ is
the charging energy of the molecule.
Our work focuses on the limit $U\to\infty$, which may be relevant when 
considering the case of  C$_{60}$ transistors whose charge states are 
most likely zero or one\cite{hparketal00,dresselhausetal96}.
$H_{leads}$ and $H_t$ are given by
\begin{equation}
H_{leads}=\sum_{i=L,R}H_{lead-i}=\sum_iH_{kin-i}+H_{int-i},
\end{equation}
and 
\begin{equation}
H_t=\sum_{k,\s\atop i=L,R}\square{t_{i}(a^{\dagger}_{i,k,\s}
+b^{\dagger}_{i,k,\s})c_{\s}+h.c.},
\end{equation}
where
\begin{multline}
H_{kin-i}=v_F\sum_{k,\s}
\left[(k-k_F)a^\dag_{k,\s,i}a_{k,\s,i}\right.\\
\left.-(k+k_F)b^\dag_{k,\s,i}b_{k,\s,i}\right]
\end{multline}
\begin{multline}
H_{int-i}=\\\frac{1}{2L_s}\sum_{q,k,k',\s,\s'}
\left\{2g_2a^\dag_{k+q,\s,i}a_{k,\s,i}
b^\dag_{k'-q,\s',i}b_{k',\s',i}\right.\\
\left.+g_4\left[a^\dag_{k+q,\s,i}a_{k,\s,i}
a^\dag_{k'-q,\s',i}a_{k',\s',i}\right.\right.\\
\left.\left.+b^\dag_{k-q,\s,i}b_{k,\s,i}
b^\dag_{k'+q,\s',i}b_{k',\s',i}\right]\right\}.
\end{multline}
$H_{kin-i}$ and $H_{int-i}$ respectively correspond 
to the kinetic and the interaction
terms of the electrons in lead $i$; $L_s$ is the linear size of the leads,
and $v_F$ ($k_F$) is the Fermi velocity (wavevector). 
Here $a^{\dagger}_{k,\s,i}(b^{\dagger}_{k,\s,i})$ creates
a right(left)-moving electron with momentum $k$ and spin $\s$ on lead $i$.
$g_2$ and $g_4$ represent forward scatterings; in our work, we will
not consider the back-scattering interaction.

In the absence of back-scattering, 
the Luttinger Hamiltonian, $H_{lead-i}$, is exactly soluble using the 
technique of bosonization\cite{kaneetal92,vondelftetal98,schulz95}. 
First, we rewrite the Hamiltonian using the Fourier
components of the particle density operator for right and left movers:
\begin{eqnarray}
\ro_{+,\s,i}(q)&=&\sum_ka^\dag_{k+q,\s,i}a_{k,\s,i},\\
\ro_{-,\s,i}(q)&=&\sum_kb^\dag_{k+q,\s,i}b_{k,\s,i}.
\end{eqnarray}
In the continuum limit ($L_s\to\infty$),
\begin{multline}
H_{kin-i}=v_F\int_0^\infty\sum_{\s}dq
\left[\ro_{+,\s,i}(q)\ro_{+,\s,i}(-q)\right.\\
\left.+\ro_{-,\s,i}(-q)\ro_{-,\s,i}(q)\right],
\end{multline}
\begin{multline}
H_{int-i}=\frac{1}{2\pi}\int dq\sum_{\s,\s'}
\left\{g_2\ro_{+,\s,i}(q)\ro_{-,\s',i}(-q)\right.\\
\left.+\frac{g_4}{2}\left[\ro_{+,\s,i}(q)\ro_{+,\s',i}(-q)
+\ro_{-,\s,i}(-q)\ro_{-,\s',i}(q)\right]\right\}.
\end{multline}
In order to express the Hamiltonian in diagonal form, we introduce the
canonically conjugate Boson operators, $\phi_{\s,i}$ and $\Pi_{\s,i}$, 
\begin{multline}
\phi_{\s,i}(x)=\\
-\frac i2\int_{q\ne 0}\frac{dq}{q}e^{-\frac{\alpha |q|}{2}-iqx}
\square{\ro_{+,\s,i}(q)+\ro_{-,\s,i}(q)}\\
\end{multline}
and
\begin{multline}
\Pi_{\s,i}(x)=\\
\frac {1}{2\pi}\int_{q\ne 0}dqe^{-\frac{\alpha |q|}{2}-iqx}
\square{\ro_{+,\s,i}(q)-\ro_{-,\s,i}(q)},
\end{multline}
where $\alpha$ is a small convergence factor or the inverse of
a large ultraviolet cutoff.
These operators are transformed into the charge and spin boson operators,
\begin{eqnarray}
\phi_{c(s),i}&=&\frac{\phi_{\uparrow,i}+(-)\phi_{\downarrow,i}}{\sqrt{2}},
\label{phi}\\
\Pi_{c(s),i}&=&\frac{\Pi_{\uparrow,i}+(-)\Pi_{\downarrow,i}}{\sqrt{2}},
\label{Pi}
\end{eqnarray}
which obey 
\begin{equation}
\square{\phi_{\mi,i}(x),\Pi_{\n,i}(y)}=i\de_{\mi,\n}\de (x-y).
\end{equation}
The single fermion operators for right- and left-moving electrons with spin
$\s$ on lead $i$ can be written in position representation as
\begin{multline}
\psi_{\pm,\s,i}(x)=\\
\lim_{\al\to 0}\frac{1}{\sqrt{2\pi\al}}e^{\pm ik_Fx-i\frac{1}{\sqrt{2}}
[\pm(\phi_{c,i}+\s\phi_{s,i})+(\thi_{c,i}+\s\thi_{s,i})]},
\end{multline}
where
\begin{equation}
\thi_{\mi,i}=\pi\int^xdx'\Pi_{\mi,i}(x').
\end{equation}
The Hamiltonian, $H_{lead-i}$, can be expressed in terms of the
canonically conjugate charge and spin boson operators (Eqs.\ref{phi},\ref{Pi}) 
as,
\begin{multline}
H_{lead-i}=\\\sum_{\mi=c,s}\frac{v_\mi}{2}\int dx
\square{\pi K_\mi\Pi^2_{\mi,i}+\frac{1}{\pi K_\mi}(\partial_x\phi_{\mi,i})^2},
\end{multline}
where
\begin{gather}
v_c=\sqrt{\round{v_F+\frac{g_4}{\pi}}^2-\round{\frac{g_2}{\pi}}^2};\\
K_c=\sqrt{\frac{\pi v_F+g_4-g_2}{\pi v_F+g_4+g_2}};\\
v_s=v_F;\\
K_s=1.
\end{gather}
Notice that $K_c  < 1$ ($K_c >1$) for repulsive (attractive) interaction. 

Now a convenient canonical transformation\cite{mahan2} can be applied on 
the full Hamiltonian making it diagonal in the dot variables. 
The desired transformation is $H\to H'=e^{s}He^{-s}$ where
\begin{equation}
s\equiv\la\sum_\s c^{\dagger}_{\s}c_{\s}\round{b^{\dagger}-b}=-s^\dag.
\end{equation}
Then, the transformed Hamiltonian is given by
\begin{equation}
H'=H'_{dot}+H_{leads}+H'_t,
\end{equation}
where
\begin{multline}
H'_{dot}=\e'\sum_\s c^{\dagger}_{\s}c_{\s}+\frac {U'}{2}\sum_\s c^{\dagger}_{\s}c_{\s}
\round{\sum_\s c^{\dagger}_{\s}c_{\s}-1}\\+\w_0b'^{\dagger}b',
\end{multline}
and
\begin{equation}
H'_t=\sum_{k,\s\atop i=L,R}\square{t_{i}e^{-\la(b^{\dagger}-b)}(a^{\dagger}_{i,k,\s}
+b^{\dagger}_{i,k,\s})c_{\s}+h.c.}.
\end{equation}
The transformed phonon operator is 
\begin{equation}
b'=b-\la\sum_\s c^{\dagger}_{\s}c_{\s},
\end{equation}
which shows that ground state energy of the phonon spectrum depends on 
electron occupancy of the dot. The shifted energy level of the molecule 
is $\e'=\e_0-\la^2\w_0$, while the shifted charging energy, $U'$, is
not explicitly shown since we only consider the limit $U\to\infty$.

\subsection{Formalism}
\subsubsection{The high-temperature approximation}
\label{hitapprox}
The spirit of the high-temperature approximation stems from the presence of 
two competing energy scales: the temperature, $k_BT$, and the tunneling rate
of an electron from a lead onto the dot (or vice versa), $h\G$. In the
approximation, one assumes that $k_BT\gg h\G$. Low tunneling rate infers that
the number of charges on the dot is a well-defined integer, and thus one can
characterize the state of the dot by a set of dot occupational probabilities.
The equation of motion of these dot occupational probabilities is the 
rate equation. In fact, a dot-lead system can be tuned into the low tunneling 
regime in certain experiments\cite{hparketal00}.

\subsubsection{The rate equation}
\label{rateeqn}

The central element in determining the transport properties of our molecular
transistor in the high-temperature approximation is the semi-classical 
rate equation, 
which are the equations of motion for the various occupational probabilities 
of the dot. To obtain the rate equation, we use the density matrix formalism, 
developed by Mitra {\it et al.}\cite{mitraetal04}, which allows us to determine
probabilities for various states of the dot under both equilibrium 
and out-of-equilibrium conditions. 

We start with the full density matrix $\ro$ that obeys the equation of motion
\begin{equation}
\dot{\ro}=-i[H,\ro].
\label{eomro}
\end{equation}
The initial step is to assume that the leads are in equilibrium 
independent of the molecular state. To harness this assumption, we 
express $\ro$ as the sum of the projected density matrix,
\begin{equation}
\ro_s=\mbox{Tr}_{leads}\{\ro(t)\}\otimes\ro_{leads},
\label{ros}
\end{equation}
and the complementary density matrix 
\begin{equation}
\ro_t=\ro-\ro_s.
\label{rot}
\end{equation} 
Here, $\ro_{leads}$ is the density matrix
of the two leads at thermal equilibrium with chemical potential
$\mi_L$ for the left lead and $\mi_R$ for the right lead. 
$\mbox{Tr}_{leads}$ denotes tracing over the
leads degree of freedom. On physical grounds, this decomposition 
is useful because the diagonal 
components of the reduced density matrix, $\ro_D\equiv\mbox{Tr}_{leads}\{\ro_s\}$, 
relate directly to the occupational probabilities of various 
states of the dot. In the high-temperature approximation, 
the equation of motion for $\ro_s$, using Eqs.\ref{eomro},\ref{ros}, can be
written as
\begin{equation}
\dot{\ro}_{sI}(t)\approx
                 -i\mbox{Tr}_{leads}[H_{tI}(t),\ro_{tI}(t)]\otimes\ro_{leads},
\label{eomros}
\end{equation}
where subscript $I$ indicates that the operators are in the interaction picture:
$O_I(t)=e^{iH_0t}O(t)e^{-iH_0t}$, where $H_0\equiv H_D+H_{leads}$. 
Eq.\ref{eomros} indicates that the time evolution of $\ro_s$ is coupled to 
$\ro_t$ and vice versa. However, one can show that the equations
of motion for $\rho_s$ and $\rho_t$ can be decoupled in the high temperature
approximation\cite{mitraetal04}, leading to
\begin{multline}
\frac{d\ro_s(t)}{dt}=\frac {d(\ro_D\otimes\ro_{leads})}{dt}
              \approx\\-\frac 12\int_{-\infty}^\infty d\ta
               \square{H_t,e^{-iH_0(t-\ta)}[H_t,\ro_s(t)]e^{iH_0(t-\ta)}}.
\label{eomrosfin}
\end{multline}
By tracing out the leads degree of freedom on both sides of Eq.\ref{eomrosfin},
one obtains coupled equations of motion for the various
occupational probabilities of the dot. Identifying $P^n_q$ as the probability
for the dot to be in a state with $n$ electrons and $q$ phonons,
\begin{equation}
P^n_q=\ang{n,q|\ro_D|n,q},
\end{equation}
we arrive at the rate equation,
\begin{equation}\begin{split}
\dot{P}^n_q=
&\sum_{i=L,R\atop p}2n_i\round{(q-p)\w_0+U(n-1)}\G^{i}_{q,p}P^{n-1}_p\\
&+\bar{n}_i\round{(p-q)\w_0+Un}\G^{i}_{q,p}P^{n+1}_p\\
&-\bar{n}_i\round{(q-p)\w_0+U(n-1)}\G^{i}_{p,q}P^n_q\\
&-2n_i\round{(p-q)\w_0+Un}\G^{i}_{p,q}P^n_q,
\end{split}\label{rate}\end{equation}
where
\begin{equation}
\G^{i}_{q<p}=t_{i}^2 N_0 \left|\sum_{l=0}^q
               \frac{(\la^2)^l\sqrt{q!p!}\la^{|q-p|}e^{-\la^2/2}}
                    {l!(q-l)!(l+|p-q|)!}\right|^2,
\end{equation}
\begin{equation}\begin{split}
n_i(\w)&=\frac{1}{N_0}\int_{-\infty}^{\infty}dte^{i\w t}
        \ang{\psi^\dag_{i,\s}(0)\psi_{i,\s}(t)}\\
       &=\frac{e^{\frac{-(\e'+\w-\mi_i)}{2T}}}{2\p}
        \round{\frac{2\p T\al}{v_c}}^{\be -1}\times\\
&\quad\times\frac{\left|\G\round{\frac\be 2+\frac{i(\e'+\w-\mi_i)}{2\p T}}\right|^2}
        {\G(\be)},
\end{split}\end{equation}
\begin{equation}\begin{split}
\bar{n}_i(\w)&=\frac{1}{N_0}\int_{-\infty}^{\infty}dte^{i\w t}
              \ang{\psi_{i,\s}(t)\psi^\dag_{i,\s}(0)}\\
             &=\frac{e^{\frac{\e'+\w-\mi_i}{2T}}}{2\p}
              \round{\frac{2\p T\al}{v_c}}^{\be -1}\times\\
 &\quad\times\frac{\left|\G\round{\frac\be 2+\frac{i(\e'+\w-\mi_i)}{2\p T}}\right|^2}
              {\G(\be)},
\end{split}\end{equation}
where $N_0 = (v_c v_F)^{-1/2}$.
$\G^{i}_{q,p}$ denotes the rate at which an electron on the dot hops off onto
lead $i$ or an electron on lead $i$ hops onto the dot while 
the dot phonon occupancy changes from $q$ to $p$ during the process; it 
is symmetric under the interchange of $q$ and $p$. $\G(z)$ is the gamma
function, while $n(\w)$ is the electron occupation number for interacting 
electrons and is analogous to the Fermi
distribution, $f(\w)$ for non-interacting electrons 
($\bar{n}(\w)$ is analogous to $(1-f(\w))$). 
The factors of 2 in the first and the 
last terms in Eq.\ref{rate} are due to electron's spin degree of freedom. 
The exponent $\be$ is given by 
\begin{equation}
\be=\frac 14\round{K_c+\frac {1}{K_c}}+\frac 12.
\end{equation}
In the non-interacting limit, $K_c=1$ and hence $\be=1$. However, for
{\it any} non-vanishing interaction, $K_c\ne 1$ and hence $\be>1$.

The general form of the rate equation is the same as the original rate 
equation for the case of non-interacting leads\cite{mitraetal04}. However, 
notice that when the interaction in the leads is taken into account, 
the Fermi functions, $f(\w)$, in the original rate equation, is 
replaced by the Luttinger-liquid distribution function, $n(\w)$. 

In the limit $U\to\infty$, $n$ in Eq.\ref{rate} is either 0 or 1.
During numerical calculations, we allow $q$ to take on values 
between 0 and some large cut-off, $q_{max}\gg 1$. In this case, $P^n_q$ 
is a vector of length $2q_{max}$, and the rate equation can be expressed in
matrix form,
\begin{equation}
\dot{\mathbf{P}}=\mathcal{M}\mathbf{P},
\label{ratematrix}
\end{equation}
where $\mathcal{M}$ is a $2q_{max}\times 2q_{max}$ matrix. 
At steady state, $\dot{\mathbf{P}}=0$, so the solution for $\mathbf{P}$ 
is the eigenvector
corresponding to the zero eigenvalue of $\mathcal{M}$. For 
equilibrated phonons, we use the ansatz\cite{mitraetal04},
\begin{equation}
P^n_q=P^ne^{-q\w_0/T}(1-e^{-\w_0/T}).
\label{Peqm}
\end{equation}
where $P^0$ is the probability that the dot is empty and $P^1=1-P^0$.

\subsubsection{I-V characteristics}

In the density matrix formalism, current through lead $i$ is given by
\begin{equation}
\ang{I_i}=\mbox{Tr}\{\ro(t)\hat{I}_i\}=\mbox{Tr}\{\ro_t(t)\hat{I}_i\},
\label{Igen}
\end{equation}
where the current operator is 
\begin{equation}
\hat{I}_i=it_{i}\sum_{k,\s}
    \square{e^{-\la(b^{\dagger}-b)}(a^{\dagger}_{i,k,\s}
    + b^{\dagger}_{i,k,\s}) c_{\s}-h.c.}.
\end{equation}
Performing the trace over the leads and the dot degrees of freedom, 
we can express the steady state current in terms of 
dot occupational probabilities, $P^n_q$. In the limit $U\to\infty$, we get
\begin{multline}
\ang{I_i}=\sum_{q,p}\left[2P^0_qn_i((p-q)\w_0)\G^i_{q,p}\right.\\
                             \left.-P^1_q\bar{n}_i((q-p)\w_0)\G^i_{p,q}\right].
\label{I}
\end{multline}

\subsubsection{DC noise characteristics}

It is also interesting to see how the second moment in current 
is affected by the electron-phonon coupling and electron-electron
interactions in the leads.
Current through the molecular transistor fluctuates even when the system is
subject to a DC bias. These current fluctuations, or noise, can be 
characterized by the zero frequency component of the 
Fourier transform $\tilde{S}(\w)$ of the current correlation function,
\begin{eqnarray}
S_{LL}(t)&=&\frac 12\ang{\square{\de I_L(t),\de I_L(0)}_+}\nonumber\\
&=&\frac 12\mbox{Tr}\{\ro(I_L(t)I_L(0)+I_L(0)I_L(t))\}\nonumber\\
&&-\square{\mbox{Tr}\{\ro I_L\}}^2,
\label{sll}
\end{eqnarray} 
where $\de I_i(t)=I_i(t)-\ang{I_i(t)}$. Charge conservation implies $I_L=-I_R$. 
Therefore, the current fluctuations are equal at both leads, giving
$S_{LL}(t)=S_{RR}(t)$. Therefore, it suffices to compute the current 
correlation only at the left lead. 

In the Heisenberg picture, current evolves over time as per 
$I_L(t)=e^{iHt}I_L(0)e^{-iHt}$. Substituting this into Eq.\ref{sll}, we get
\begin{equation}
S_{LL}(t)=\frac 12\mbox{Tr}\{R(t)I_L(0)\},
\label{sllr}
\end{equation}
where 
\begin{equation}
R(t)=e^{-iHt}\square{\de I_L(0)\ro+\ro\de I_L(0)}e^{iHt}.
\end{equation}
Since $S_{LL}(t)=S_{LL}(-t)$, $S_{LL}(t)$ will be computed only for
positive time. The equation of motion for the 
causal function $R(t)$ can be obtained from the equation of motion for
the density matrix and is given by
\begin{multline}
\dot{R}(t)=\\-i[H,R(t)]+\de (t)[\de I_L(0)\ro+\ro\de I_L(0)],\quad t\ge 0,
\label{eomr}
\end{multline}
and is zero elsewhere. Solving the equation of motion for $R(t)$ 
is similar to the procedure for solving the density matrix equation of motion
(section \ref{rateeqn}). First, we
decompose the causal function into the diagonal component
in the dot and leads variables, and the off-diagonal component in those
variables: $R=R_s+R_t$, where 
$R_s=R_D\otimes R_{leads}$ with $R_D \equiv {\rm Tr}_{leads} \{R_s \}$. From Eq.\ref{eomr},
we obtain the equation of motion for $R_s(t)$ in the high-temperature 
approximation,
\begin{equation}
\dot{R}_s(t)\approx-i[H_t,R_t]+\de (t)\round{I\ro_t+\ro_tI-2\ang{I}\ro_s}.
\label{eomrs}
\end{equation}
Notice that the time evolution of $R_s$ couples to $R_t$.
The coupled equations of motion for $R_s$ and $R_t$ can be
decoupled in the high temperature approximation.
Then, as was done for the density matrix, we trace out the leads degree of freedom
in the decoupled equation of motion for $R_s$ 
and arrive at the equation of motion for $R_D$.
\begin{equation}
\dot{\mathbf{R}}_D(t)=\mathcal{M}\mathbf{R}_D+\de (t)\mathbf{h},
\label{eomrd}
\end{equation}
where $\mathcal{M}$ is the same matrix as the one appearing in 
Eq.\ref{ratematrix} and vector $\mathbf{h}$ is given by
\begin{multline}
h^n_q=-2\ang{I_L}P^n_q\\+2\sum_{p}\left[(3-n)P^{n-1}_p
                                     n_L((q-p)\w_0+U(n-1))\G^L_{q,p}\right.\\
      \left.-(n+1)P^{n+1}_p\bar{n}_L((p-q)\w_0+Un)\G^L_{q,p}\right].
\label{h}
\end{multline}

The final step is to express the DC noise in terms of dot occupational 
probabilities and the components of $\mathbf{R}_D$. 
We first rewrite Eq.\ref{sllr} as
\begin{equation}
S_{LL}(t)=\frac 12\mbox{Tr}\{R_t(t)I_L(0)\}.
\end{equation}
The trace over the dot and the leads degrees of freedom can be
performed by following the procedure used for the density 
matrix calculation.
The resulting expression can be Fourier transformed via,
\begin{equation}
\tilde{S}_{LL}(\w=0)=2\int_{-\infty}^{\infty}dtS_{LL}(t),
\end{equation}
and we arrive at the zero frequency current noise ($U\to\infty$),
\begin{multline}
\tilde{S}_{LL}(\w=0)=
       2\sum_{q,p}\left[2(\tilde{R}(0)_q^0+P^0_q)n_L((q-p)\w_0)\G^L_{q,p}\right.\\
       \left.-(\tilde{R}(0)^1_q-P^1_q)\bar{n}_L((q-p)\w_0)\G^L_{p,q}\right],
\end{multline}
where
$\tilde{\mathbf{R}}(0)$ is the Fourier transform of vector $\mathbf{R}_D(t)$ at 
$\omega=0$  and $\mathbf{R}_D(t)$ is given by
\begin{equation}
R^n_q(t)=\ang{n,q|R_D(t)|n,q}.
\end{equation}
Here $\tilde{\mathbf{R}}(0)$ can be obtained, at steady state, via Eq.\ref{eomrd}.
\begin{equation}
\tilde{\mathbf{R}}(0)=-\mathcal{M}^{-1}\mathbf{h}.
\label{R}
\end{equation}

\begin{widetext}

\section{Results}
\label{results}
\subsection{I-V characteristics}
\label{current}

The tunneling current is obtained at steady state for both equilibrated and
non-equilibrated phonons. Initially, the dot occupational probability vector, 
$P^n_q$, is numerically obtained via the rate equation (Eq.\ref{ratematrix}). 
For equilibrated phonons, the ansatz, Eq.\ref{Peqm}, is used. 
The current is then numerically computed using Eq.\ref{I}.

The current is plotted for three different electron-phonon coupling constants
$\la=0.5,1.0,2.0$. For each $\la$, current is computed under symmetric bias
($\mi_L=-\mi_R=V_{sd}/2$) and completely asymmetric bias ($\mi_R=0$, $\mi_L=V_{sd}$) 
conditions and both equilibrated and unequilibrated phonon distributions 
are considered. 
Each plot consists of five plot-lines with different line styles. Each line style 
corresponds to a particular value of the exponent, $\be$, in the Luttinger-liquid
distribution function. The correspondence is shown in Fig.\ref{fig:current1}(a). 
All other similar plots follow the same convention. All current plots are produced
at $T=0.05\w_0$. 
\begin{figure}[t]
\begin{center}
\includegraphics[scale=0.7]{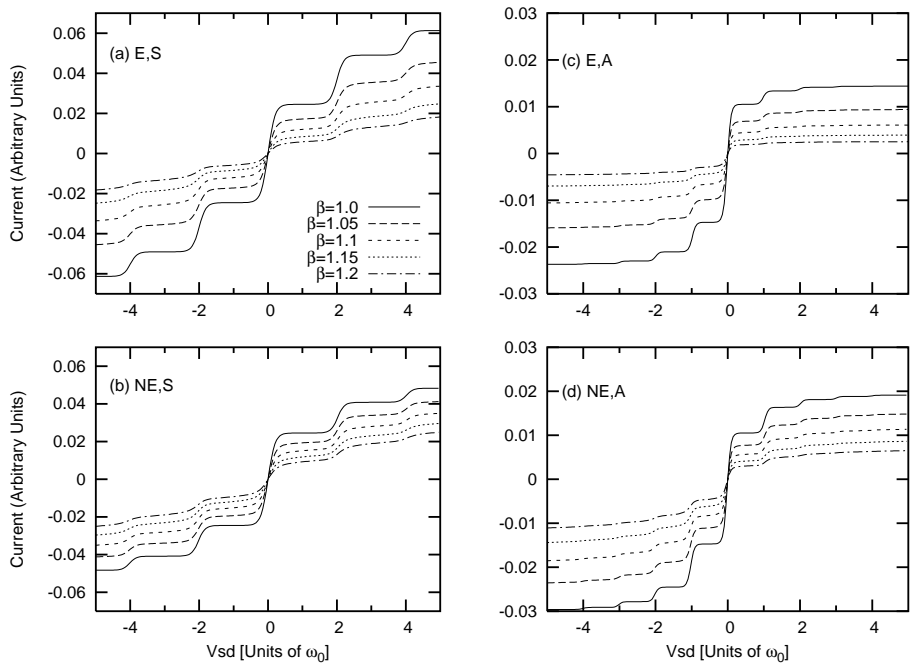}
\caption{\label{fig:current1} Tunneling current plotted as a function of
source-drain voltage with $\la=1.0$. E and NE stand 
for equilibrated and non-equilibrated phonons respectively, and 
S and A stand for symmetric and asymmetric biases respectively. $\be$ is the
exponent in the Luttinger-liquid distribution, which, in the noninteracting limit, is 1
and bigger than 1 for both repulsive and attractive interactions.}
\end{center}
\end{figure}
\begin{figure}[t]
\begin{center}
\includegraphics[scale=0.7]{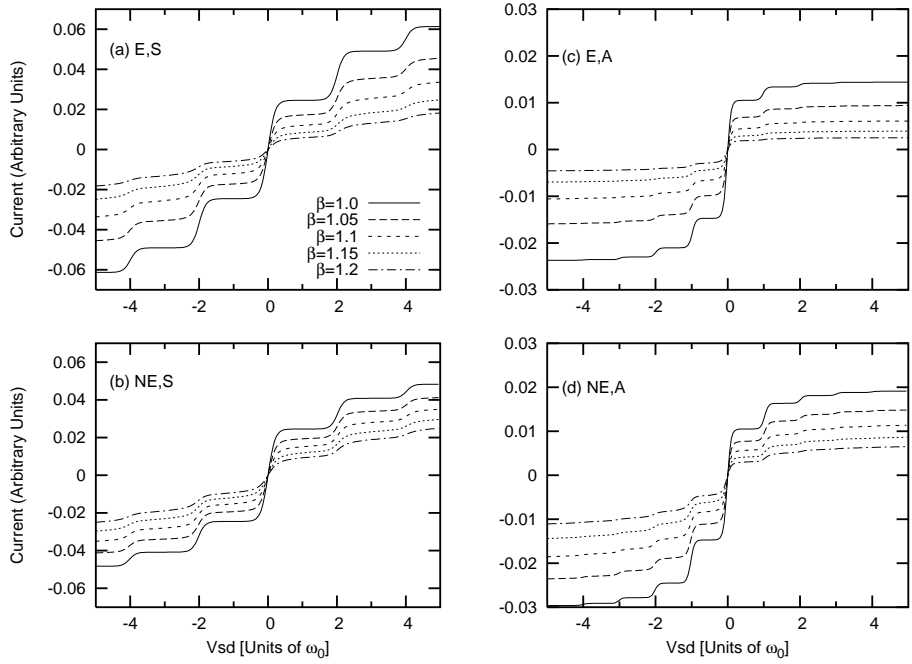}
\caption{\label{fig:current2} Tunneling current plotted as a function of
source-drain voltage with $\la=0.5$.}
\end{center}
\end{figure}
\begin{figure}[h]
\begin{center}
\includegraphics[scale=0.7]{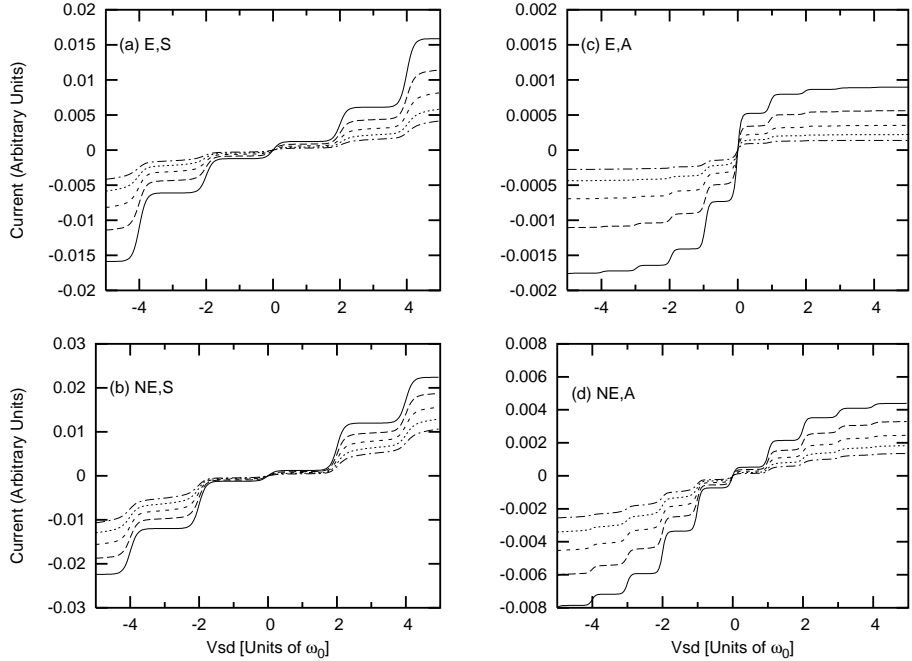}
\caption{\label{fig:current3} Tunneling current plotted as a function of
source-drain voltage with $\la=2.0$.}
\end{center}
\end{figure}
\begin{figure}[t]
\begin{center}
\includegraphics[scale=0.5,angle=270]{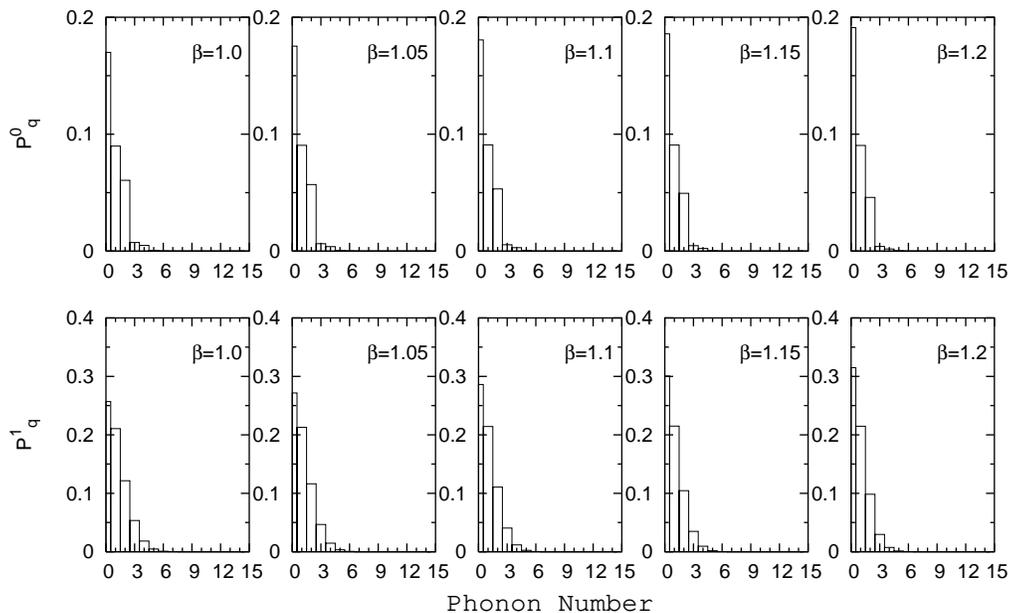}
\caption{\label{fig:phdist1} Phonon probability distribution plotted 
with $\la=1.0$ in the case of non-equilibrium phonons. Here, 
$\mi_L=-\mi_R=2\w_0$.}
\end{center}
\end{figure}
\begin{figure}[t]
\begin{center}
\includegraphics[scale=0.7]{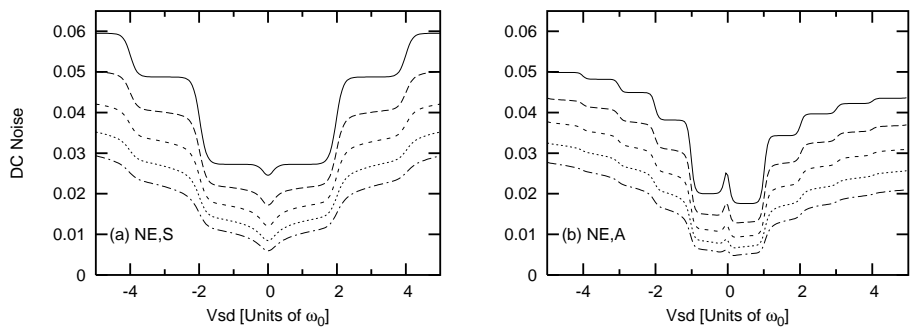}
\caption{\label{fig:noise1} DC noise plotted as a function of
source-drain voltage with $\la=1.0$.}
\end{center}
\end{figure}
\begin{figure}[t]
\begin{center}
\includegraphics[scale=0.7]{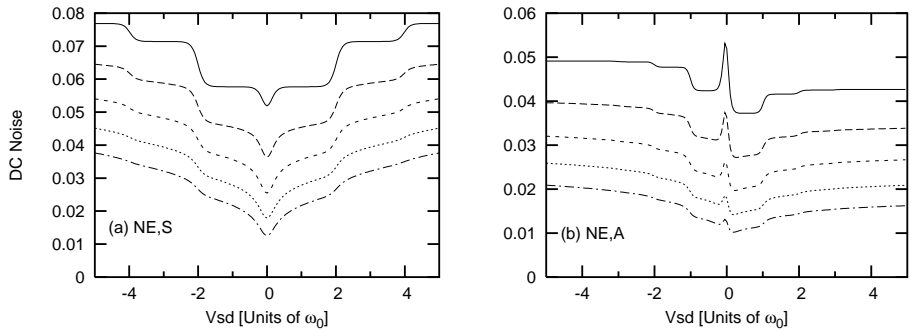}
\caption{\label{fig:noise2} DC noise plotted as a function of
source-drain voltage with $\la=0.5$.}
\end{center}
\end{figure}
\begin{figure}[t]
\begin{center}
\includegraphics[scale=0.7]{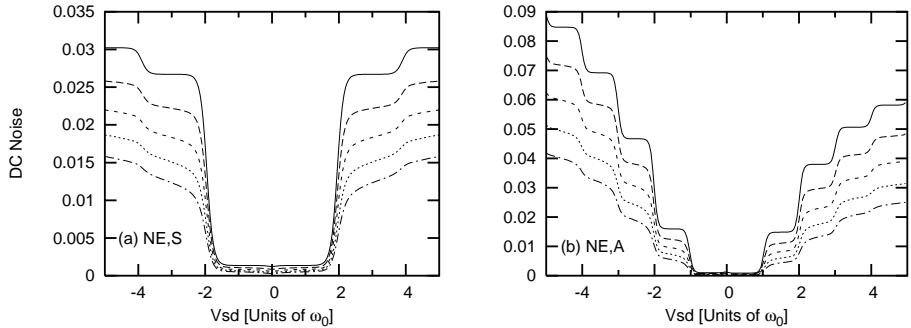}
\caption{\label{fig:noise3} DC noise plotted as a function of
source-drain voltage with $\la=2.0$.}
\end{center}
\end{figure}
\begin{figure}[t]
\begin{center}
\includegraphics[scale=0.7]{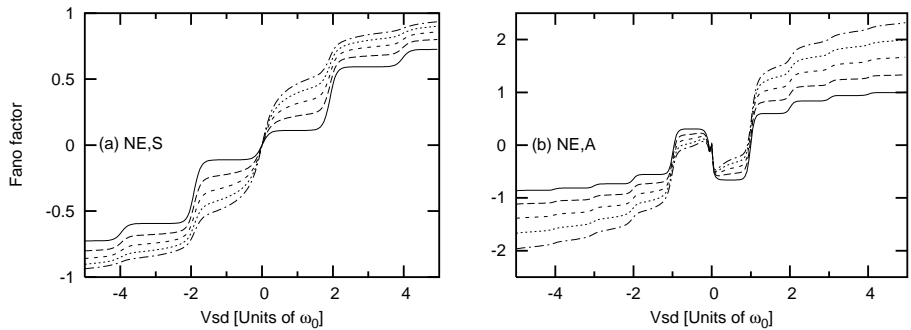}
\caption{\label{fig:fano1} Fano factor plotted as a function of
source-drain voltage with $\la=1.0$.}
\end{center}
\end{figure}
\begin{figure}[t]
\begin{center}
\includegraphics[scale=0.7]{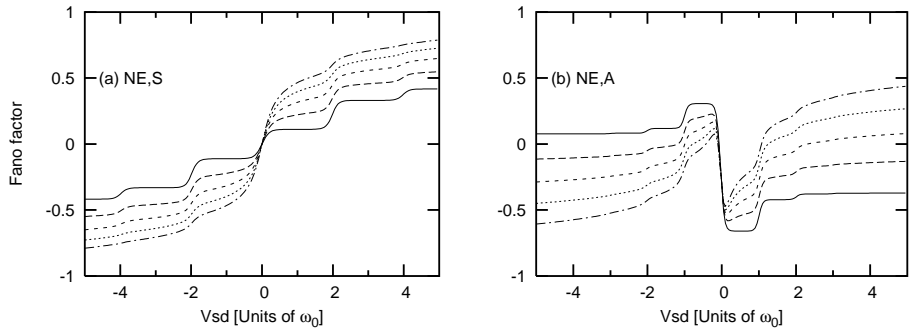}
\caption{\label{fig:fano2} Fano factor plotted as a function of
source-drain voltage with $\la=0.5$.}
\end{center}
\end{figure}
\begin{figure}[t]
\begin{center}
\includegraphics[scale=0.7]{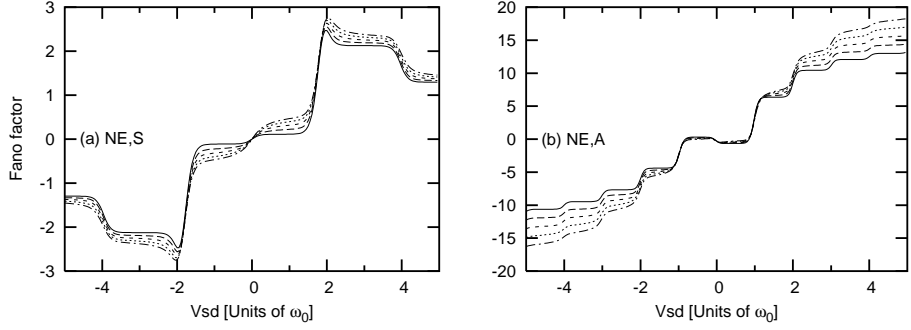}
\caption{\label{fig:fano3} Fano factor plotted as a function of
source-drain voltage with $\la=2.0$.}
\end{center}
\end{figure}
\begin{figure}[h]
\begin{center}
\includegraphics[scale=0.75]{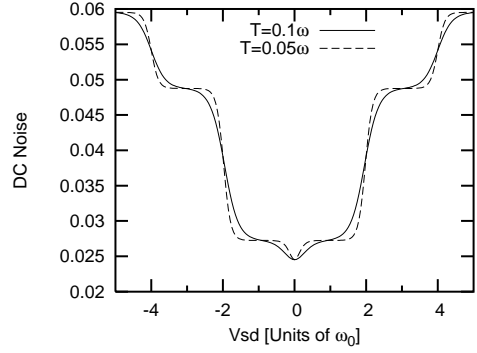}
\caption{\label{fig:noise4} DC noise plotted as a function of
source-drain voltage with $\la=1.0$ and $\be=1$ at two
different temperatures.}
\end{center}
\end{figure}

We first compare currents for equilibrated and non-equilibrated phonons. 
Mitra {\it et al}.\cite{mitraetal04} pointed out a peculiar behaviour 
in the current; near $\la=1.0$ and with symmetric bias, the current is larger
for equilibrated phonons than for non-equilibrated phonons. This is
an unusual result since one would expect that phonons would arrange 
themselves so as to maximize the current in the non-equilibrium case. 
They attribute this peculiarity to the choice of $\la$ at which many of the
higher-order diagonal matrix elements (corresponding to 
$q$-phonon-$q$-phonon processes), $\G^i_{qq}$, are suppressed. Although
their work involved non-interacting leads, one can
see in Fig.\ref{fig:current1} that this behaviour is observed also in the 
case of interacting leads. 
The behaviour, however, is not observed when $\la=2$, consistent with 
Mitra {\it et al}.'s claim that this anomalous behaviour is pertinent to
$\la$ values close to 1. 

The most apparent consequence of introducing interaction in the leads
is the suppression of the tunneling current; the current
decreases as interaction strength is raised. 
The question is whether the suppression of the current is mainly due to
changes in the phonon distribution induced by electron interaction
in the leads or a more direct consequence of the Luttinger-liquid
distribution function.
Fig.\ref{fig:phdist1} plots the phonon probability
distribution for $\la=1.0$.
Upon careful inspection, a slight narrowing in the phonon distribution
is observed; probabilities of dot states with low phonon numbers are
increased while the wing of the distribution, corresponding to the
dot states with high phonon numbers, is suppressed. Although they are not shown, 
the same subtle change in the phonon distribution is observed for 
other values of the electron-phonon coupling. However, because these 
changes are very small, the suppression in the current is unlikely due 
to the changes in the phonon distribution. 
Thus the main effect comes from 
the Luttinger-liquid distribution function that has replaced 
the Fermi distribution in the expression for the current (Eq.\ref{I}). 
The Luttinger-liquid distribution function represents
single-particle-like excitations decaying faster over time 
than non-interacting electrons. This results
in the suppression of the overall current.

Although Fig.\ref{fig:phdist1} shows that the phonon distribution 
is not a strong function of electron interaction in the
leads, Mitra {\it et al.}\cite{mitraetal04} showed that the phonon
distribution is strongly affected by the electron-phonon coupling, $\la$. 
In particular, the distribution was farther from equilibrium for weak coupling 
than for strong coupling. We have observed that the phonon distribution strongly depends
on the electron-phonon coupling even in the presence of intralead electron
interaction.
We conclude that varying the electron-phonon 
coupling has a much larger effect on the phonon distribution than varying the 
interaction in the leads.

Another noticeable effect on the tunneling current due to interaction in
the leads is the smearing and change in the slopes of the plateaus. 
These plateaus, which occur in a range of bias where the tunneling electrons
encounter no new accessible excited state of the molecule,
are flat in the case of non-interacting leads. However, as the interaction
increases, the slopes also increase. This observation can be explained 
qualitatively in the zero-temperature limit, 
which is not the appropriate limit for this work but may give insight 
into this issue. The single-particle time correlation function for
the Luttinger-liquid behaves as $1/t^{\be}$. 
Fourier transform of this correlation function, 
which relates directly to $n(\w)$, then behaves as 
$\w^{\be-1}$. Therefore, the current should show 
$V_{sd}^{\be-1}$-type behaviour at low voltages. 
Each time you excite a new phonon, current should show 
$(V_{sd}-V_0)^{\be-1}$-type behaviour, where $V_0$ is the voltage at
which a new phonon is excited. Indeed, in the non-interacting limit, where
$\be=1$, the slope is zero. However, when $\be>1$, the tilt and the smearing
of the plateaus result.


\subsection{DC noise characteristics}
DC noise characteristics are obtained under both symmetric and asymmetric
bias conditions for non-equilibrated phonons. The noise for equilibrated
phonons is not presented here. In section \ref{current}, we argued that the
modification in the current due to interaction in the leads is mainly
due to the Luttinger-liquid nature of the electrons in the leads, not so
much the induced change in the phonon distribution. For this reason, we
anticipate that the effects of interaction on the DC noise for 
equilibrated phonons will be very similar for the case of non-equilibrated
phonons. 

The noise is plotted at the same three values of $\la$. 
The correspondence between the plot-line styles and the interaction 
parameters follow the same convention as in the plots for the current (see Fig.1).
Figs.\ref{fig:noise1}-\ref{fig:noise3} show that the noise
is suppressed by interaction in the leads. This is expected since
interaction in the leads causes tunneling events to be more correlated.
The increase in correlation among subsequent tunneling events leads to 
decreased fluctuations and hence decreased current noise.

In Figs.\ref{fig:fano1}-\ref{fig:fano3}, we plot the current-normalized
DC noise, or the Fano factor $F$,
\begin{equation}
F(V_{sd})=\frac{S(V_{sd})-S(V_{sd}=0)}{2eI(V_{sd})}.
\end{equation}
For all values of $\la$, electron interaction in the leads results
in an increase in $F$. The plots also show that $F$ depends dramatically
on $\la$. In particular, $F$ grows rapidly as $\la$ increases \cite{oppen04},
and is found to be super-Poissonian for $\lambda > 1$ (see Fig.\ref{fig:fano3}). 

Fig.\ref{fig:noise4} plots the noise at a higher temperature, $T=0.1\w_0$. 
Raising the temperature has a similar qualitative effect on the
noise as increasing the interaction in the leads;
the steps in the noise curve are
smeared out and the plateaus between steps acquire a slope. However,
the overall noise amplitude does not change when the temperature is raised
while the amplitude decreases when the interaction is introduced.
\vspace{0.5cm}
\end{widetext}

\section{Conclusion}
\label{conclusion}
In this paper, we investigated the steady state current and the DC 
noise characteristics of a single level
quantum dot coupled to two Luttinger liquid leads. In particular, we
studied how the electron interaction in the leads and the coupling 
between tunneling electrons and on-dot
phonons affect the dot's transport properties. We considered both equilibrium
and out-of-equilibrium phonon distributions, and both symmetric and
asymmetric bias orientations. The density matrix formalism was used to harness 
both phonon distributions in our calculations and to derive the semi-classical
rate equation for the dot occupational probabilities. The calculations were done in 
the high-temperature approximation, in which the electron tunneling rate 
was assumed to be small enough so that the dot charge fluctuations are 
insignificant.

We have found that interaction in the leads in general 
suppresses the overall tunneling current and the noise, and 
smears out the steps in the
$I$-$V$ characteristics. This effect is mainly a consequence of 
the Luttinger-liquid correlation in the leads. The change in
the phonon distribution of the dot due to electron-electron interaction
in the leads is relatively small albeit we can see a slight narrowing of the
phonon distribution as the interaction strength is increased, leading
to only a minor contribution to the changes in current and noise.
Comparing these behaviours of the current and noise with the 
effect of increased temperature, we find that increasing the temperature
does not suppress overall current and noise even though it
smears out the steps in $I$-$V$ and noise characteristics.
It is also found that the Fano factor, that represents the noise normalized
by the current, becomes enhanced as the intralead electron interaction
gets bigger. The Fano factor increases rapidly as 
the electron-phonon interaction becomes stronger and, interestingly,
shows super-Poissonian behaviour as the electron-phonon 
coupling becomes greater than order one.

{\bf Acknowledgement}: This work was supported by the NSERC of Canada, 
Canadian Institute for Advanced Research, Canada Research Chair Program,
and OGS.
We would like to thank Eugene Kim for earlier collaboration. 


\end{document}